\documentclass[aps,prx,twocolumn,showpacs,superscriptaddress,longbibliography]{revtex4-2}
\usepackage{amsmath, amsfonts, amssymb,graphicx}
\usepackage{graphicx,epstopdf}
\usepackage{gensymb}
\usepackage[dvipsnames]{xcolor}
\newcommand{\be}{\begin{equation}}
\newcommand{\ee}{\end{equation}}
\newcommand{\bea}{\begin{eqnarray}}
\newcommand{\eea}{\end{eqnarray}}
\newcommand{\bse}{\begin{subequations}}
\newcommand{\ese}{\end{subequations}}

\usepackage{multibib}
\usepackage{color}
\usepackage[colorlinks,bookmarks=false,citecolor=darkblue,linkcolor=red,urlcolor=blue]{hyperref}

\definecolor{darkred}{rgb}{0.7,0.0,0.0}

\definecolor{darkblue}{rgb}{0,0.02,0.45}

\definecolor{darkgreen}{rgb}{0.02,0.45,0.0}

\definecolor{violet}{rgb}{0.8,0.2,0.6}

\begin{document}

\preprint{APS/123-QED}

\title{Evidence of random spin-singlet state in a three-dimensional quantum spin liquid candidate Sr$_3$CuNb$_2$O$_9$}

\author{S. M. Hossain}
\affiliation{Department of Physics, Shiv Nadar Institution of Eminence, Gautam Buddha Nagar, Uttar Pradesh 201314, India}
 
\author{S. S. Rahaman}
\address{S.N. Bose National Centre for Basic Sciences, Salt Lake, Kolkata 700106, India}

\author{H. Gujrati}
\affiliation{Department of Physics, Shiv Nadar Institution of Eminence, Gautam Buddha Nagar, Uttar Pradesh 201314, India}

\author{Dilip Bhoi}
\address{The Institute for Solid State Physics, University of Tokyo, Kashiwa, Chiba 277-8581, Japan.}

\author{A. Matsuo}
\address{The Institute for Solid State Physics, University of Tokyo, Kashiwa, Chiba 277-8581, Japan.}

\author{K. Kindo}
\address{The Institute for Solid State Physics, University of Tokyo, Kashiwa, Chiba 277-8581, Japan.}

\author{M. Kumar}
\email{manoranjan.kumar@bose.res.in}
\affiliation{S.N. Bose National Centre for Basic Sciences, Salt Lake, Kolkata 700106, India}

\author{M. Majumder}
\email{mayukh.cu@gmail.com}
\affiliation{Department of Physics, Shiv Nadar Institution of Eminence, Gautam Buddha Nagar, Uttar Pradesh 201314, India}

\date{\today}           

\begin{abstract}
Disorder is ubiquitous in any quantum many-body system and is usually considered to be an obstacle to the elucidation of the underlying physics of complex systems, but its presence can often introduce exotic phases of matter that cannot generally be realized in a clean system. We report here a detailed experimental  and theoretical  study of magnetic properties of highly disordered Sr$_3$CuNb$_2$O$_9$ material  which exhibits random site mixing between Cu and Nb. The magnetic moments (Cu$^{2+}$) are arranged in a quasi-cubic (three-dimensional) manner, leading to a high degree of frustration with a Curie-Weiss temperature ($\theta_{CW}$) of about -60 K without any long-range magnetic ordering down to 466 mK. These observations suggest that Sr$_3$CuNb$_2$O$_9$ is a candidate for a quantum spin liquid. More interestingly, the susceptibility ($\chi = M/\mu_0H$) and the $C_m/T$ ($C_m$ is the magnetic part of the heat capacity) follow a power-law behavior with decreasing temperature. In addition, $M(T,\mu_0H)$ and $C_m(T,\mu_0H)/T$ show scaling relationships over a wide temperature and field range. This unusual behavior with respect to the conventional behavior of a QSL can be discussed qualitatively as the coexistence of a disorder-induced random spin singlet (RSS) state  and a QSL state. A quantitative description has been given by numerical calculations considering a power-law probability distribution $P(J) \propto J^{-\gamma}$ ($J$ is the exchange interaction) of random spin singlets. The parameters extracted from the numerical calculations are in excellent agreement with the experimental data. Furthermore, the analytical results are also consistent with the power-law and scaling behavior of $\chi$ and $C_m(T,\mu_0H)/T$ as a whole. Thus, our comprehensive experimental and theoretical analysis provides evidence for the stabilization of the RSS state in a three-dimensional lattice. 
\end{abstract}
                            
\maketitle

\textit{Introduction:} The presence of disorder is inevitable in any real-life 
quantum many-body system and it hinders us from elucidating the actual physics of a system. More specifically, the effects of disorder on the phases or phase transitions in quantum many-body systems is reflected by the disappearance of spontaneous symmetry breaking or smearing out the singularities associated with phase transitions and critical phases \cite{doi:10.1146/annurev-conmatphys-031218-013433}. On the contrary, there are also examples of exotic ground states and new phenomena driven by disorder, which can not be realized in a clean or disorder-free system. Thus, discovering and characterizing those ground states in a quantum many-body system where there exists the interplay of disorder and quantum fluctuations is a current field of study in condensed matter physics \cite{doi:10.1146/annurev-conmatphys-031218-013433, doi:10.1146/annurev-conmatphys-033117-053925}. These systems are also crucial for several application purposes, e.g. they can be the key elements for memories and state transfer channels in quantum computing \citep{doi:10.1146/annurev-conmatphys-031214-014726, PhysRevLett.106.040505, Smith2016, KITAEV20062}. One of the first prominent examples of disorder-driven states is Anderson localization, where the wavefunction of non-interacting quantum particles is highly localized near a point in space due to a strong random potential \cite{PhysRev.109.1492, 10.1063/1.3206091}. Recently, disorder-driven topological Anderson localization has also been observed  \cite{Lin2022, BHATT2021168438}. Quasi-particle interference (QPI) is an powerful experimental tool and it originates because of the presence of disorder and helps to explore the momentum space informations in a 2D system\cite{https://doi.org/10.1002/adma.201707628}. Disorder-driven phenomena are also often observed in correlated many-body systems, e.g. Kondo disorder state, quantum Griffiths phase, which shows exotic non-Fermi liquid behaviors \cite{doi:10.1146/annurev-conmatphys-031218-013433, doi:10.1146/annurev-conmatphys-031016-025531, Shender1996, Bergman2007}. Experimentally and theoretically, it was also observed that the presence of a strong disorder can turn the first-order magnetic phase transition to a second-order phase transition near the quantum critical point, and thus disorder-induced quantum critical point can be obtained \cite{RevModPhys.88.025006, doi:10.1146/annurev-conmatphys-033117-053925}.

The effects of disorder in frustrated magnets is nowadays a focus area. Frustration may lead to a huge accidental degeneracy of different spin configurations and the ground state is the superposition of the degenerate states, known as the quantum spin liquid (QSL) state. In the QSL state, due to the absence of spontaneous symmetry breaking, the spins are dynamic even at T = 0~K and one can expect exotic fractionalized excitations.\cite{ANDERSON1973153, Balents2010, doi:10.1126/science.aay0668}. The presence of  disorder in a frustrated system gives rise to a glassy state, the spin glass (SG) state \cite{RevModPhys.58.801}. Super-lattice structure produced by site disorder may also give rise to a frustrated triangular lattice and further may lead to the observation of a QSL state\cite{PhysRevLett.125.267202}. Very recently, it has been pointed out by Kimchi \textit{et al.} that in some disordered frustrated systems, peculiar features in heat capacity and magnetization can be qualitatively explained as the presence of a $"\text{random spin-singlet}"$ (RSS) state admixed with QSL state \cite{Kimchi2018}. The RSS state have been first discussed for doped semiconductors in which the antiferromagnetic exchange energies follow a power-law probability distribution due to the presence of disorder \cite{PhysRevB.22.1305, PhysRevB.34.387, ROY1986513}. Further, this idea of RSS state has also been extended to two-dimensional (2D) and three-dimensions (3D) \cite{PhysRevLett.48.344, PhysRevLett.61.597, PhysRevLett.74.2543, PhysRevB.37.5531} for doped semiconductors. Whereas, only a handful of the 2D frustrated systems show such exotic states where RSS state and QSL coexist \cite{PhysRevB.106.174406, PhysRevLett.125.117206, PhysRevResearch.2.013099, PhysRevB.101.020406} and so far the presence of such a state has not been observed in any 3D frustrated system experimentally. Note that in some 3D pyrochlore systems, the signature of QSL-like state was interpreted to be due to a RSS state but the probability distribution of exchange energies was not a power-law probability distribution \cite{PhysRevLett.123.087201}. Thus, discovering new three-dimensional systems with strong frustration along with disorder and elucidating the complex ground state qualitatively and quantitatively is a primary goal in this field. Needless to say that stabilizing such a state in a 3D lattice is inherently more challenging compared to low-dimensional systems, primarily due to the enhanced dimensionality reduces quantum fluctuations and hence more tendency to order magnetically.  

In this letter, we have reported a detailed study of a three-dimensional quasi-cubic system Sr$_3$CuNb$_2$O$_9$. Magnetization and heat capacity measurements down to 466~mK indicate the absence of long-range magnetic ordering (LRO) despite a strong exchange interactions (of the order of -60~K) between Cu$^{2+}$-moments and thus established Sr$_3$CuNb$_2$O$_9$ as a 3D QSL candidate. Furthermore, the power-law divergence of susceptibility and heat capacity, along with the scaling behaviors of magnetization and heat capacity, indicate the presence of RSS state admixed with QSL due to the presence of strong Cu/Nb site disorder. Furthermore, numerical and analytical calculations described the experimental data quantitatively.

%%%%%%%%%%%%%%%%%%%%%%%%%%%%%%%%%%%%%%%%%%%%%%%%%%%%%%%%%%%%%%%%%%%%%%%%%%%%%%
\begin{figure}[h]
       \centering
       \includegraphics[width=0.5\textwidth]{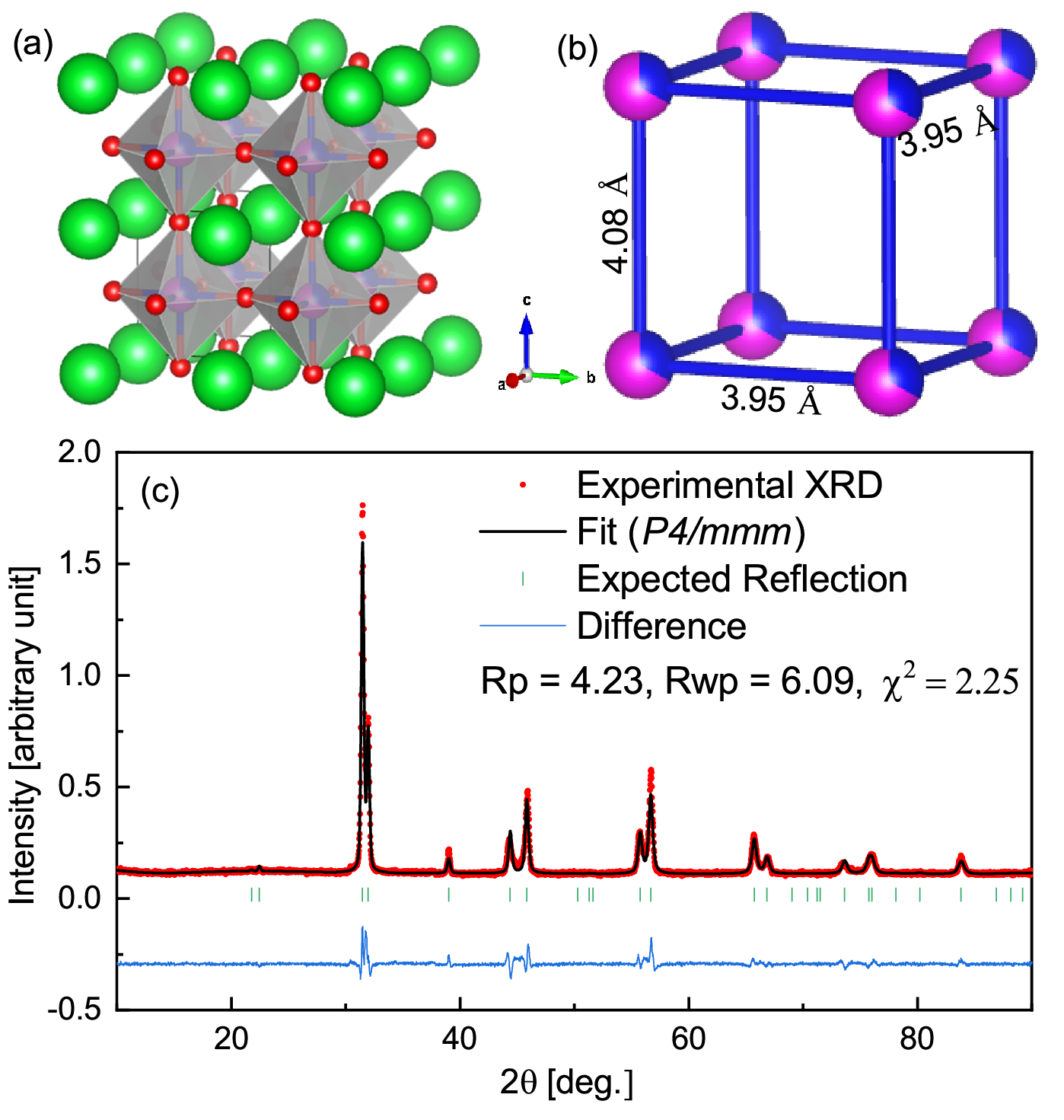}
       \caption{(a) Magnetic sites forming an octahedral environment with the oxygen atoms (red) and the green spheres showing strontium atoms intercalated in between. (b) Arrangement of the magnetic Cu$^{2+}$ (blue) sharing the same site with the non-magnetic Nb$^{5+}$ (magenta).(c) A clean XRD pattern shows the absence of any impurity and has been analysed with the Rietveld refinement method.}
       \label{fig: XRD pattern}
\end{figure}
%%%%%%%%%%%%%%%%%%%%%%%%%%%%%%%%%%%%%%%%%%%%%%%%%%%%%%%%%%%%%%%%%%%%%%%%%%%%%%

\textit{Crystal structure:} 
%Very recently, Sr$_3$CuSb$_2$O$_9$ have been studied in detail. The Cu and Sb share the same site, but occupying the same site is not random but rather periodic (shows up as a super-lattice peak in the XRD pattern), and that leads to the frustrated edge-shared triangular lattice of Cu, which is then responsible for a QSL ground state. 
Here Sr$_3$CuNb$_2$O$_9$ has been synthesized via a conventional solid-state reaction technique (details have been discussed in the supplementary materials\cite{Supp}). Sr$_3$CuNb$_2$O$_9$ belongs to a family of triple perovskite crystal structures with a general formula A$_3$B$_3$O$_9$ shown in Fig.\ref{fig: XRD pattern}(a). The compound stabilises in a tetragonal crystal structure of space group P4/mmm (space group no. 123), with lattice parameters ($a=b\approx$ 3.953~Å), $c$ = 4.089~Å, and $\alpha$ = $\beta$ = $\gamma$ = 90$^{\circ}$ obtained from PXRD rietveld refinement Fig.\ref{fig: XRD pattern}(c). No extra peaks represent the existence of single-phase formation. Most importantly, no super lattice peaks have been detected, indicating a random distribution of Cu and Nb atoms in contrast to Sr$_3$CuSb$_2$O$_9$~\cite{PhysRevLett.125.267202}. Thus, the magnetic skeleton is formed by Cu atoms randomly distributed over Nb sites, with occupancy ratios of 1/3 and 2/3, respectively and the in-plane distance between magnetic ions $\approx$ 3.95~Å, whereas out-of-plane distance is $\approx$ 4.07~Å, makes it almost a cubic magnetic lattice structure shown in Fig.\ref{fig: XRD pattern}(b). Such a quasi-cubic 3D structure, if one considers the next nearest-neighbour interactions along with nearest-neighbour, may give rise to a 3D frustrated lattice. Thus, the present compound on the one hand, may provide a 3D frustrated lattice and, on the other hand, provide strong disorder (random site-disorder of Cu and Nb), which may then be favourable to realize a RSS state. In order to verify our expectation from the structural analysis, we have explored the ground state by performing detailed magnetization and heat capacity measurements in a broad temperature-field region. 

%%%%%%%%%%%%%%%%%%%%%%%%%%%%%%%%%%%%%%%%%%%%%%%%%%%%%%%%%%%%%%%%%%%%%%%%%%%%%%
\begin{figure*}
       \centering
       \includegraphics[width=1\textwidth]{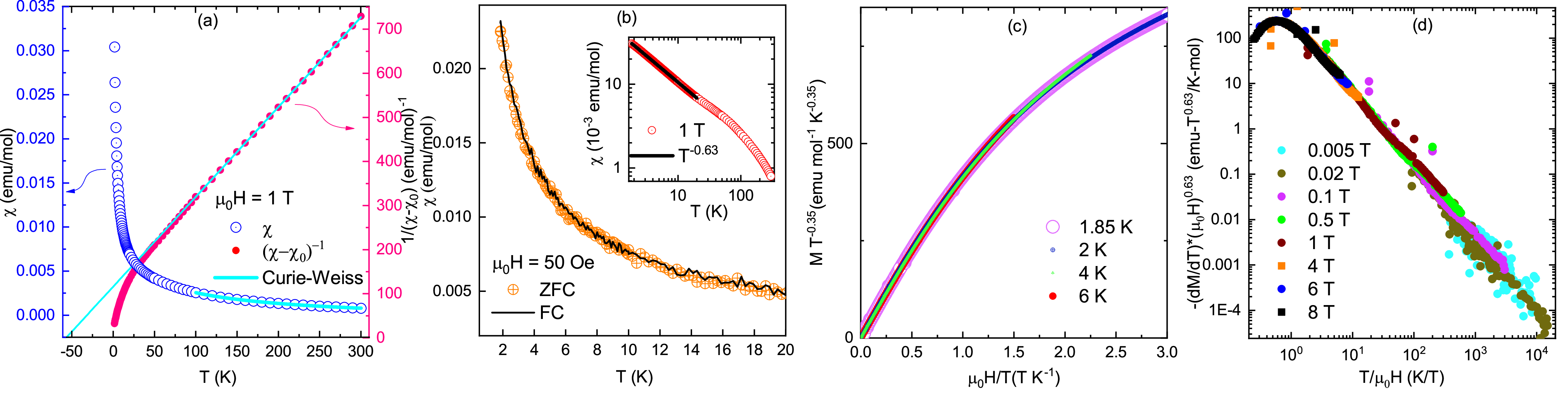}
       \caption{(a) $\chi$ versus temperature at an applied magnetic field of 1~T (ZFC) and extrapolation of the Curie-Wiess fit (ranging from 100-300 K) in the inverse susceptibility data. (b) ZFC and FC data at 50 Oe applied magnetic field, and the inset shows the power-law behavior with the exponent $\gamma=0.63$. (c) Scaling behavior of the magnetization data with respect to field at different temperatures with an exponent of $\gamma-1=0.35$. (d) Scaling of the magnetization with respect to the temperature is showing excellent data collapse in a wide applied magnetic field range from 50 Oe to 8 T.}
       \label{fig: Magnetic Susceptibility}
\end{figure*} 
%%%%%%%%%%%%%%%%%%%%%%%%%%%%%%%%%%%%%%%%%%%%%%%%%%%%%%%%%%%%%%%%%%%%%%%%%%%%%%

\textit{Magnetization:} Magnetic susceptibility ($\chi=M/\mu_0H$) has been measured as a function of temperature at different applied magnetic fields. No anomaly associated with long-range magnetic ordering (LRO) has been observed down to 1.85 K as seen from Fig.\ref{fig: Magnetic Susceptibility}(a). The susceptibility data between 100-300~K, measurement at an applied  magnetic field of 1~T, has been fitted with Curie-Weiss law ($\chi=\frac{C}{T-\theta_{CW}}+\chi_0$) shown as a solid line in Fig.\ref{fig: Magnetic Susceptibility}(a). Curie-Weiss temperature ($\theta_{CW}$) obtained $\approx$ -60 K with a curie constant $C=0.49$ cm$^3$Kmol$^{-1}$. The negative sign indicates the nature of exchanges is antiferromagnetic with a high degree of frustration parameter, defined as $f=\theta_{CW}/T_N$. For any magnetic system, $\theta_{CW}$ indicates the strength of exchange interactions between magnetic moments, and $T_N$ is the ordering temperature. In the present systems, even though the moments are highly correlated with high $\theta_{CW}$, the system cannot go to an LRO state at least down to 1.85~K due to strong frustration. The effective moment has been calculated using the formula $\mu_{eff}=\sqrt{8C}$ gives 1.98 $\mu_B$, where $C$ is the Curie constant obtained from $1/{\chi}$ vs $T$ fitting shown in Fig. \ref{fig: Magnetic Susceptibility}(a). This indicates that the magnetic ions are in Cu$^{2+}$ (S=1/2) state. Furthermore, the absence of branching in $\chi$ between the zero-field cooling and field cooling measured at 50 Oe rules out the possibility of spin freezing state (e.g. spin glass) at least down to 1.85~K (Fig. \ref{fig: Magnetic Susceptibility}(b)). Thus, magnetization measurements ensure that Sr$_3$CuNb$_2$O$_9$ can be considered as a highly frustrated magnetic system promising to have a QSL ground state. 

More interestingly, if one closely observes the temperature dependence of $\chi$, then one can see that $\chi$ follows a power-law behavior ($T^{-\gamma}$) between 20~K to 1.85~K with $\gamma$ to be 0.63. Such a power-law behavior was predicted to be a signature of RSS state in a frustrated system\cite{Kimchi2018}. Along with the power-law in $\chi$, the $M$ versus $\mu_0H$ data at different temperatures also collapses in one curve following a scaling behavior as shown in Fig.\ref{fig: Magnetic Susceptibility}(c) as $MT^{\gamma-1}$ against $\mu_{0}H/T$. Even a scaling has also been observed in $(dM/dT)\times(\mu_{0}H)^\gamma$ versus $T/\mu_{0}H$. All these power-law and scaling behaviors support the presence of the predicted RSS state admixed with a QSL state.  

\begin{figure*}
       \centering
       \includegraphics[width=0.95\textwidth]{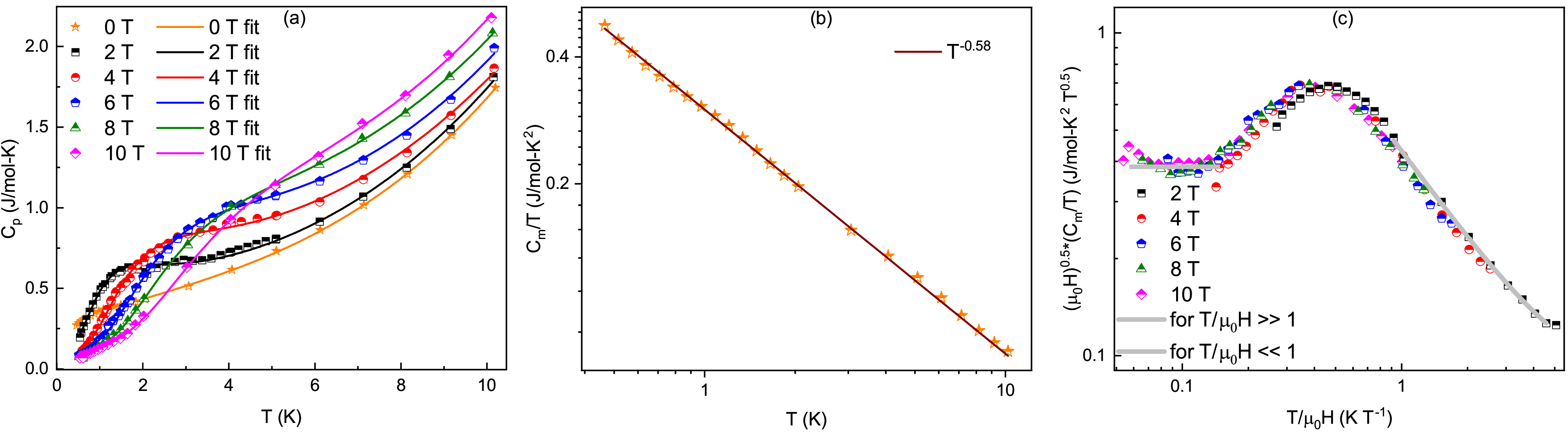}
       \caption{(a) Temperature dependence of $C(T,\mu_0H)$ at different applied magnetic fields. Solid lines show the fit considering three different contributions described in the text. (b) Power-law fit of the $C_m/T$ (at 0~T) as a function of temperature with the an exponent of 0.58. (c) Scaling of the $C_m(T,\mu_0H)/T$.}
       \label{fig: Low temp heat capacity}
\end{figure*}

\textit{Heat capacity:} Heat capacity ($C_p(T,\mu_0H)$) measurements have been performed from 300~K down to 466~mK at various applied magnetic fields to explore the ground state of Sr$_3$CuNb$_2$O$_9$. Fig.\ref{fig: Low temp heat capacity}(a) shows $C_p$ versus $T$ at different applied magnetic fields. The absence of any lambda-like anomaly indicates that Sr$_3$CuNb$_2$O$_9$ is avoided of any LRO down to 466~mK even though there is a strong correlation between magnetic moments and thus established that Sr$_3$CuNb$_2$O$_9$ is a 3D QSL candidate with a high degree of frustration ($f=\theta_{CW}/T_{min}\simeq 120$, where $T_{min}$ represents the lowest temperature heat capacity has been measured). In general, $C_p/T$ follows either exponential behaviour or power-law ($T^\gamma$) for a gapped and gapless QSL, respectively\cite{PhysRevB.88.220413, PhysRevB.96.174411}. In contrast to this expectation, $C_p/T$ follows a power-law behavior of $T^{-\gamma}$ in the present system at zero applied magnetic field. To get an qualitative description of the $C_p$ data, we modeled it by considering three contributions as
\begin{equation}
C_p(T,\mu_0H)=C_{ph}+C_{Sch}+C_{power-law}
\end{equation}
where $C_{power-law}=A_{power-law}T^{1-\gamma}$ \cite{PhysRevB.106.174406}. The other two terms $C_{ph}$, $C_{Sch}$ represent phonon and two-level Schottky contribution (representing the hump), respectively. At sufficiently low temperatures, the phonon contribution behaves as $C_{ph}=\beta_{ph}T^3$, where $\beta_{ph}$ is the Debye coefficient. And
\begin{equation}
C_{Sch} = A_{Sch}R\left[\frac{\Delta(\mu_0H)}{k_{B}T}\right]^2\frac{e^{\left(\frac{\Delta(\mu_0H)}{k_B T}\right)}}{\left[1 +  e^{\left(\frac{\Delta(\mu_0H)}{k_B T}\right)}\right]^2}
\end{equation}
where $\frac{\Delta}{k_B}$ represents the energy gap. 

%%%%%%%%%%%%%%%%%%%%%%%%%%%%%%%%%%%%%%%%%%%%%%%%%%%%%%%%%%%%%%%%%%%%%%%%%%%%%% 
%%%%%%%%%%%%%%%%%%%%%%%%%%%%%%%%%%%%%%%%%%%%%%%%%%%%%%%%%%%%%%%%%%%%%%%%%%%%%%

$C_m(\mu_0 H = 0)/T$ at zero field has been plotted as a function of temperature in Fig.\ref{fig: Low temp heat capacity}(b), which was estimated by subtracting the phonon (C$_{ph}$) and Schottky contributions (due to 1.8\% of weakly coupled impurities; for details see \cite{Supp}) from the total $C_p$. $C_m(\mu_0 H = 0)/T$ shows a power-law divergence with an exponent $\gamma = 0.58$ with lowering temperature. Also, note that the exponent $\gamma$ is in agreement with the obtained $\gamma$ from magnetization measurements. 

%The finite $\Delta$ at zero applied field indicates the presence of weakly coupled impurity spins. The fraction was estimated to be about 1.8\% (details are mentioned in the supplementary material). The observation of a small weakly coupled impurity contribution at zero field is a common feature of several other QSL candidates\cite{PhysRevB.106.174406,PhysRevLett.125.117206,Fujihala2020}. If this fraction is field independent then they represent the impurity contributions\cite{PhysRevLett.125.267202,PhysRevLett.125.117206}, otherwise they originates from the main part of the sample. To have a very good fit of the $C_p$ versus temperature data at different fields, we had to vary the fraction of the Schottky contribution which is found to be increasing with increasing applied magnetic field (almost saturates to 35\% at 10~T). Thus indicates that at finite applied magnetic field, $C_{Sch}$ is s part of the total magnetic contribution to $C_p$ and hence $C_m = C_{power-law}+C_{Sch}$ as has been seen elsewhere\cite{}??. $C_m/T$ at zero field has been plotted as a function of temperature in Fig.\ref{fig: Low temp heat capacity}(b), which was estimated by subtracting the phonon (C$_{ph}$) and Schottky contributions (C$_{Sch}$: due to weakly coupled impurities) from the total $C_p$. $C_m/T$ shows a power-law divergence with an exponent $\gamma = 0.58$ with lowering temperature. Also, note that the exponent $\gamma$ is in agreement to the exponent $\gamma$ from magnetization measurements. 

From the qualitative analysis of the $C_p$ it is now clear the spins are correlated (as it is seen from high $\theta_{CW}$) and may form random singlets by a power-law distribution of exchange interactions due to the presence of strong site-disorder. Because of the power-law distribution of the exchange interactions (responsible for forming singlets), the excitations are gapless overall, giving rise to the power-law behavior in magnetization or heat capacity and eventually the scaling behaviors. These correlated spins (which form the singlets) may also form resonating state due to the presence of strong frustration, which may give rise to a QSL behavior. 

%Furthermore, the presence of a power-law distribution of exchange interactions also assures a fraction of spins are coupled by a vanishingly small exchange interactions. As a result, they can be considered as nearly isolated orphan spins, and those spins contribute as the $C_{Sch}$ term (estimated to be around 2\%) in the total heat capacity. 

Kimchi \textit{et al.}\cite{Kimchi2018} pointed out that the magnetic contribution of the total heat capacity at finite fields (represented by $C_m  (\mu_0H) = C_{p}-C_{ph}$; see  \cite{Supp} for details) for this disorder-induced state, collapses into a single curve of the form
\begin{equation}
\frac{{C_m[\mu_0H,T]}}{T}\sim \frac{1}{{(\mu_0H)^\gamma }}F_q\left[ {\frac{T}{\mu_0H}} \right],
\end{equation}
where $Fq(X)$ is a scaling function which is determined by the energy distribution of the singlets
\begin{equation}
F_q[X]\sim \left\{ {\begin{array}{*{20}{c}} {X^q} & {X \ll 1} \\ {X^{ - \gamma }\left( {1 + \frac{{c_0}}{{X^2}}} \right)} & {X \gg 1} \end{array}} \right.
\end{equation}
where $q=1$ and $q=0$ represent the situation with and without Dzyaloshinskii-Moriya (DM) interactions in the effective low-energy theory coupling the orphan spins, respectively. Fig.\ref{fig: Low temp heat capacity}(b) represents $(\mu_0H)^{0.5} \times C_m/T$ as a function of $T/\mu_0H$ in which all the $C_m$ data points fall on top of each other and thus validates the scaling relationship as proposed by the theory. In the range where $X<<1$, the data saturates, indicating the absence of DM interactions, whereas in the limit of $X>>1$, the data agrees reasonably well with the predicted scaling function Fig.\ref{fig: Low temp heat capacity}(c). It should be pointed out that, the scaling behavior still holds for the total heat capacity (see Fig.6 of the supplementary material\cite{Supp}) except for the high-temperature region where the lattice contribution dominates. 

%This further illustrates that apart from the lattice contribution to the $C_p$, the rest of the $C_p$ at finite applied magnetic field is the total magnetic contribution ($C_m = C_{power-law}+C_{Sch}$).  

Thus, the detailed magnetization and heat capacity measurements in a broad temperature and field range indicate a coexistence of RSS state and QSL state in a three-dimensional Sr$_3$CuNb$_2$O$_9$. Even though the experimental data represented here qualitatively matches the theoretical prediction, more quantitative description will be crucial to elucidate the underlying complex physics. 

\textit{Theory:} We further carried out numerical calculations to have a quantitative description of the temperature and field dependence of 
magnetization and heat capacity to confirm the possibility of the presence of RSS state in Sr$_3$CuNb$_2$O$_9$. To understand the magnetic 
properties of the material we first analyze the percentage distribution ($\%\rho$) of Cu in various clusters of sizes up to 30x30x30 cubic 
lattice of the compound Sr$_3$CuNb$_2$O$_9$ with only $33\%$ of Cu. The distance between the two Cu atoms sitting in two nearest octahedra of the present compound
is approximately 4~Å, therefore, a spin is a part of cluster if the distance of a spin is less than equal to 4~Å from any of spin in the cluster and with this criterion 
the percentage distribution ($\%\rho$) of the Cu atoms decays exponentially with cluster size (L) as shown in the supplementary material with the details of the calculation\cite{Supp}. The most dominating spin clusters are free spins and part of these spins may be forming weakly interacting Cu structures 
with random exchange as the distances among the spins are larger than 4~Å and random. These spins can have comparable nearest, next-nearest exchanges due to their distribution in three-dimensional space. Therefore, these weakly interacting frustrated spins can be treated as random dimer singlet with a broad distribution of the exchanges. 
Other spin cluster bigger than dimer are very small in numbers and may have negligible effect on the magnetic properties of the material. Hence, we adopt an isotropic dimer Heisenberg spin-1/2 model Hamiltonian, incorporating a distribution of exchanges. This idea is inspired by the spin-1/2 model proposed by Dasgupta, Ma, and Fisher \cite{PhysRevB.22.1305} for highly disordered one-dimensional systems, as well as the model presented by Bhatt and Lee \cite{PhysRevLett.48.344} for highly disordered two and three-dimensional systems. The dimer model can be written as  
\begin{equation}
H=\sum_i J_i S_{i,1} . S_{i,2}=\sum_i H_i 
\end{equation} 
where $H_i$ is $i^{th}$ isolated dimer Hamiltonian for spin 1 and 2, and $J_i$ is isotropic exchange interaction of the $i^{th}$ dimer. We assume $J_i$ can take any continuous values (in between 0 and 1 in the reduced unit) and it follows the distribution $P(J) = J^{-\gamma}$, i.e. the smaller J has a higher probability distribution. A detailed description of the calculation is given in the Supplementary file\cite{Supp}.\\

%%%%%%%%%%%%%%%%%%%%%%%%%%%%%%%%%%%%%%%%%%%%%%%%%%%%%%%%%%%%%%%%%%%
\begin{figure}[h]
\centering
    \includegraphics[width=0.8\linewidth]{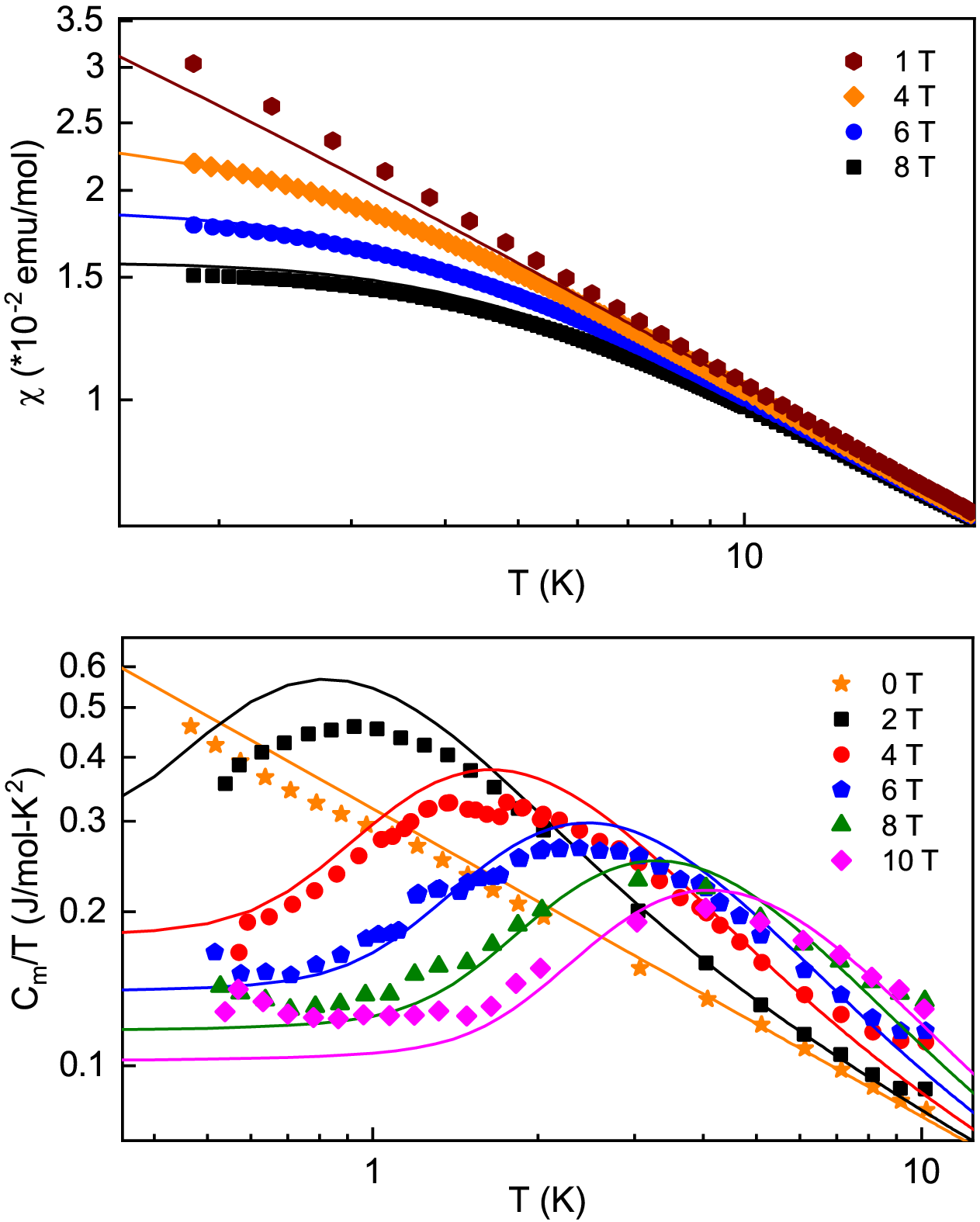}
    \caption{Upper panel: Temperature dependence of $\chi$ (obtained experimentally) represented by the scattered points and the solid lines indicate the data obtained numerically. Lower panel: The solid lines and the scattered points represents experimentally and numerically obtained data of $C_m/T$.}
    \label{fig:plot_set}
\end{figure}
%%%%%%%%%%%%%%%%%%%%%%%%%%%%%%%%%%%%%%%%%%%%%%%%%%%%%%%%%%%%%%%%%%%

We use $g=2.22$, the average $J= <J_i> \approx 72K$ and $\gamma=0.6$ to fit experimental curve of $C_m(T,\mu_0H)/T$ and $\chi(T)$. The $\frac{M}{H}$ for four different fields are shown in the upper panel of Fig. \ref{fig:plot_set}, and  $\chi (=\frac{M}{H})$ curves show a power-law decay as a function of temperature. The experimental curve of $C_m(T,\mu_0H)/T$ is shown as points and the theoretical curve as lines for six different magnetic fields, all the experimental curves can be fitted well with the theoretical curves as shown in lower panel of Fig. \ref{fig:plot_set}. In the lower panel of Fig. \ref{fig:plot_set}, the linear behavior of the $C_m/T$ plot clearly shows the power-law behavior for the zero-field but finite field data shows criticality as a function of $T$ and $H$. This behavior is well known as data-collapse in $\frac{T}{H}$ for the resonating limit $|J-H|< k_BT$. It is noticed that both  $C_m(T,\mu_0H)/T$ and $\chi$ can be fitted well with the experimental data with the same exponent which is $0.6$ in this case. We analyse these results for different limits of temperature and field in %\sout{the next section} Sec.\ref{sec:zero-field} and Sec.\ref{sec:finite-field} of 
the Supplementary file\cite{Supp}.

Thus, our numerical calculations quantitatively established that the temperature and field dependence of magnetization and heat capacity are due to the presence of disorder-driven RSS state.

\textit{Conclusion:} The present manuscript has studied a novel disorder-induced state in great detail. We report here a comprehensive experimental and theoretical investigation of the compound Sr$_3$CuNb$_2$O$_9$. The disorder introduced by site-mixing between Cu and Nb and frustration due to competing nearest-neighbour and next nearest-neighbour of the three-dimensional quasi-cubic lattice gives rise to an exotic state in which RSS state coexists with QSL. Numerical calculations further gave a quantitative description of the experimental results and provided evidence of the realization of the RSS state in a three-dimensional lattice. Analytical calculations again confirm the experimentally observed behaviors of $\chi$ and $C_m(T,\mu_0H)/T$. These observations established Sr$_3$CuNb$_2$O$_9$ as one of the rare compounds where such a RSS occurs in a 3D-lattice.

\textit{Acknowledgement:} SMH and MM would like to acknowledge UGC-DAE Collaborative Research Scheme (ref. CRS/2021-22/01/393) for funding. MK thanks DST-SERB for funding through projects CRG/2020/000754. 

%\printbibliography

\bibliography{References}
%\bibitem{}
%\textbf{page},845-566(2023)
%\blindtext
\end{document}